\newif\ifComment\Commenttrue 
\documentclass[conference]{IEEEtran}
\usepackage{microtype}
\usepackage{amsmath,amssymb,amsfonts,amsthm} 
\usepackage{comment}                         
\usepackage{ifthen}                          
\usepackage{minibox}                         
\usepackage{multirow}                        
\usepackage{suffix}                          
\usepackage{graphicx}                        
\usepackage[hyphens]{url}                    
\usepackage{xargs}                           
\usepackage[dvipsnames]{xcolor}              
\usepackage{hyphenat}                        
\usepackage{fancyvrb}                        
\usepackage{supertabular}                    
\usepackage{xspace}                          

\newcommand\ie{i.e.,\xspace}
\newcommand\eg{e.g.,\xspace}

\newcommand{\mypara}[1]{\smallskip\noindent\emph{\textbf{{#1.}}}}
\newcommand{\myparax}[1]{\smallskip\noindent\emph{\textbf{{#1}}}}

\usepackage[switch]{lineno}
\usepackage[frozencache]{minted}

\newcommand{\ccode}[1]{\mintinline[fontsize=\small]{c}{#1}}

\newminted[cblock]{c}{xleftmargin=\parindent, linenos=none, breaklines=true, fontsize=\small}
\newminted[cblockindent]{c}{xleftmargin=1em, linenos=none, breaklines=true, fontsize=\small}

\usepackage{stmaryrd}
\usepackage{booktabs}
\usepackage{array}
\usepackage{enumerate}
\usepackage[square,comma,numbers,sort&compress]{natbib}

\usepackage{tikz}
\usetikzlibrary{external}
\usepackage{letltxmacro}
\reversemarginpar
\usepackage{todonotes}
\usepackage{marginnote}
\usepackage{tikz}

\usepackage[colorlinks=true,linkcolor=Green,citecolor=NavyBlue,pagebackref]{hyperref} 
\usepackage[hyphenbreaks]{breakurl}  
\usepackage[nameinlink,capitalise,noabbrev]{cleveref}            

\newcommand\llabel[1]{\hypertarget{#1}{\linelabel{#1}}}
\renewcommand\llabel[1]{}

\newcommand{\etal}{et al.\xspace}
\newcommand{\contract}[2]{\ensuremath{\llbracket \,\cdot\, \rrbracket^{\text{#1}}_{\text{#2}}}}
\newcommand{\p}{p\xspace}
\newcommand{\contractP}[2]{\ensuremath{\llbracket \p \rrbracket^{\text{#1}}_{\text{#2}}}}
\renewcommand{\ni}[1]{\ensuremath{\mathit{NI}({#1})}}
\newcommand{\nipi}[2]{\ensuremath{\mathit{NI}({#1}, {#2})}}

\renewcommand{\implies}{\Rightarrow}

\newcommand{\satisfies}{\vdash}

\theoremstyle{definition}
\newtheorem{definition}{Definition}

\newcommand{\gb}[1]{\textcolor{blue}{GB: #1}}
\newcommand{\sj}[1]{\textcolor{orange}{SC: #1}}
\newcommand{\cd}[1]{\textcolor{magenta}{CD: #1}}
\newcommand{\dm}[1]{\textcolor{violet}{DM: #1}}

\newcommand{\NEW}[1]{{#1}}

\ifComment
\else
\renewcommand{\gb}[1]{\ignorespaces}
\renewcommand{\sj}[1]{\ignorespaces}
\renewcommand{\cd}[1]{\ignorespaces}
\renewcommand{\dm}[1]{\ignorespaces}
\renewcommand{\notedm}[1]{\ignorespaces}
\renewcommand{\notesj}[1]{\ignorespaces}
\renewcommand{\notecd}[1]{\ignorespaces}
\renewcommand{\TODO}[1]{\ignorespaces}
\fi

\begin{document}

\title{\huge{SoK: Practical Foundations for Software Spectre Defenses}}

\newcommand{\firstauthormark}{$^*$}

\author{
\IEEEauthorblockN{Sunjay Cauligi}
\IEEEauthorblockA{\textit{UC San Diego}}
\IEEEauthorblockA{\textit{MPI Security \& Privacy}}
\and
\IEEEauthorblockN{Craig Disselkoen}
\IEEEauthorblockA{\textit{UC San Diego}}
\and
\IEEEauthorblockN{Daniel Moghimi}
\IEEEauthorblockA{\textit{UC San Diego}}
\and
\IEEEauthorblockN{Gilles Barthe}
\IEEEauthorblockA{\textit{MPI Security \& Privacy}}
\IEEEauthorblockA{\textit{IMDEA Software Institute}}
\and
\IEEEauthorblockN{Deian Stefan}
\IEEEauthorblockA{\textit{UC San Diego}}
}

\maketitle


\begin{abstract}
  Spectre vulnerabilities violate our fundamental assumptions about architectural
  abstractions, allowing attackers to steal sensitive data despite previously
  state-of-the-art countermeasures.
  To defend against Spectre, developers of verification tools and
  compiler-based mitigations are forced to reason about microarchitectural
  details such as speculative execution.
  In order to aid developers with these attacks in a principled way,
  the research community has sought formal foundations for speculative execution
  upon which to rebuild provable security guarantees.

  This paper systematizes the community's current knowledge about software
  verification and mitigation for Spectre.
  We study state-of-the-art software defenses, both with and without
  associated formal models, and use a cohesive framework to compare the
  security properties each defense provides.
  We explore a wide variety of tradeoffs
  in the expressiveness of formal frameworks, the complexity of defense
  tools, and the resulting security guarantees.
  As a result of our analysis, we suggest practical choices for developers of
  analysis and mitigation tools, and we identify several open problems in
  this area to guide future work on grounded software defenses.
  %
  %
\end{abstract}

\section{Introduction}
\label{sec:intro}
Spectre attacks have upended the foundations of computer
security~\cite{Kocher2019spectre}.
With Spectre, attackers can steal secrets across security boundaries---both
hardware boundaries provided by the process abstraction~\cite{codejail}, and
software boundaries provided by memory safe languages and software-based fault
isolation (SFI) techniques~\cite{wahbe-sfi}.
%
In response, the security community has been working on program analysis tools
to both find Spectre vulnerabilities and to guide mitigations (e.g., compiler
passes) that can be used to secure programs in the presence of this class of
attacks.
But Spectre attacks---and speculative execution in general---violate our
typical assumptions and abstractions and have proven particularly challenging
to reason about and defend against.

Many existing defense mechanisms against Spectre are either incomplete (and
thus miss possible attacks) or overly conservative (and thus slow).
For example, the MSVC compiler's \verb#/Qspectre# pass---one of the first
compiler-based defenses against Spectre~\cite{mscvcspectre}---inserts
mitigations by finding Spectre gadgets (or patterns).
Since these patterns are not based on any rigorous analysis, the compiler
misses similarly vulnerable code patterns~\cite{specfuzz}.
As another example, Google Chrome adopted process isolation as its core defense
mechanism against Spectre attacks~\cite{site-isolation}.
This is also unsound: Canella \etal~\cite{Canella2019}, for example, show that
Spectre attacks can be performed across the process boundary.
%
%
On the other side of the spectrum, inserting fences at every load or control
flow point is sound but prohibitively slow~\cite{swivel}.

Language-based security can help us achieve---or at least understand
the trade-offs of giving up on---\emph{performance} and \emph{provable security
guarantees}.
Historically, the security community has turned to language-based security to
solidify intricate defense techniques---from SFI enforcement on
x86~\cite{rocksalt}, to information flow control
enforcement~\cite{sabelfeld2003ifc}, to eliminating side-channel attacks with
constant-time programming~\cite{barthe2014}.
At the core of language-based security are \emph{program semantics}---rigorous
models of program behavior which serve as the basis for \emph{formal security
policies or foundations}.
These policies help us carefully and explicitly spell out our assumptions about
the attacker's strength and ensure that our tools are sound with respect to
this class of attackers---e.g., that Spectre vulnerability-detection or
-mitigation tools find and mitigate the vulnerabilities they claim to find and
mitigate.

Formal foundations are key to performance too.
Without formalizations, Spectre defenses are usually either overly conservative
(which leads to unnecessary and slow mitigations) or crude (and thus
vulnerable).
For example, speculative load hardening~\cite{RFC} is \emph{safe}---it safely
eliminates Spectre-PHT attacks---but is overly conservative and slow: It
assumes that \emph{all} array indexing operations must be hardened.
In practice, this is not the case~\cite{vassena2021,lvioptim}.
Crude techniques like oo7~\cite{oo7} are both inefficient \emph{and}
unsafe---they impose unnecessary restrictions yet also miss vulnerable code
patterns.
Foundations allow us to craft defenses that are minimal (e.g., they target the
precise array indexes that need
hardening~\cite{vassena2021,guarnieri2021contracts}) and provably secure.

Alas, not all foundations are equally practical.
Since speculative execution breaks common assumptions about program
semantics---the cornerstone of language-based methods---existing Spectre
foundations explore different design choices, many of which have important
ramifications on defense tools and the software produced or analyzed by these
tools (\Cref{tab:mechanics}).
For instance, one key choice is the \emph{leakage model} of the semantics,
which determines what the attacker is allowed to observe.
Another choice is the \emph{execution model}, which simultaneously
captures the attacker's strength and which Spectre variants the resulting
analysis (or mitigation) tool can reason about.
These choices in turn determine which \emph{security policies} can
be verified or enforced by these tools.

While formal design decisions fundamentally impact the soundness and precision
of Spectre analysis and mitigation tools, they have not been systematically
explored by the security community.
For example, while there are many choices for a leakage model, the
constant-time~\cite{barthe2014} and sandbox
isolation~\cite{guarnieri2021contracts} models are the most pragmatic;
leakage models that only consider the data cache trade off security
for no clear benefits (\eg scalability or precision).
As another example, the most practical execution models borrow (again) from
work on constant-time: They are detailed enough to capture practical attacks,
but abstract across different hardware---and are thus useful for both
software-based verification and mitigation techniques.  Models which
capture microarchitectural details like cache structures make the analysis
unnecessarily complicated: They do not fundamentally capture additional
attacks and give up on portability.

\NEW{
\mypara{Contributions}
In this paper, we systematize the community's knowledge on Spectre foundations and
identify the different design choices made by existing work and their tradeoffs.
%
%
This complements existing, excellent
surveys~\cite{Canella2019,canella2020evolution,xiongsurvey} on the low-level
details of Spectre attacks and defenses which do not consider foundations or,
for example, high-level security policies.
}
Throughout, we discuss the limitations of existing formal frameworks, the defense
tools built on top of these foundations, and future directions for research.
%
%
%
%
%
%
In summary, we make the following contributions:
%
\begin{itemize}
  \item Study existing foundations for Spectre analysis in the form of
  semantics, discuss the different design choices which can be made in a
  semantics, and describe the tradeoffs of each choice.
  \item Compare many proposed Spectre defenses---both with and without formal
    foundations---using a unifying framework, which allows us to understand
    differences in the security guarantees they offer.
  \item Identify open research problems, both for foundations and for
  Spectre software defenses in general.
  \item Provide recommendations both for developers and for the research
    community that could result in tools with stronger security guarantees.
\end{itemize}


\mypara{Scope} \NEW{
In this systematization, we focus on software-only defenses against Spectre attacks.
We focus on \emph{Spectre} because most other transient attacks (e.g.,
Meltdown~\cite{meltdown}, LVI~\cite{vanbulck2020lvi}, MDS~\cite{Intel2019MDS},
or Foreshadow~\cite{Vanbulck2018}) can efficiently be addressed in the hardware,
through microcode updates or new hardware designs.
(This is also the reason existing software-based tools against transient
execution attacks focus solely on Spectre, as we discuss in Section~\ref{sec:otherattacks}.)
%
We focus on \emph{defenses} because prior work, notably Canella
\etal~\cite{Canella2019}, already give an excellent overview of the types of
Spectre vulnerabilities and the powerful capabilities they give attackers.
And we focus on \emph{software-only} defenses---%
although proposals for hardware defenses are extremely valuable, hardware design
cycles (and hardware upgrade cycles) are very long.
Moreover, software foundations are useful for understanding hardware and
hardware-software co-designs (\eg they directly affect execution and leakage
models).
Having secure software foundations allows us to defend against
today's attacks on today's hardware, and tomorrow's as well.
%
%
}

\section{Preliminaries}
\label{sec:prelims}
In this section, we first discuss Spectre attacks and how they violate
security in two particular application domains: high-assurance cryptography
and isolation of untrusted code.
Then, we provide an introduction to formal semantics for security and its
relevance to secure speculation in these application~domains.

\subsection{Spectre vulnerabilities}
%
Spectre~\cite{Kocher2019spectre,Maisuradze2018spectre5,Koruyeh2018spectre5,Horn2018spectre4,branch-predictors,straight,psf,kiriansky2018speculative}
is a recently discovered family of vulnerabilities
stemming from \emph{speculative execution} on modern processors.
Spectre allows attackers to learn sensitive information by causing
the processor to mispredict the targets
of control flow (\eg conditional jumps or indirect calls) or data flow
(\eg aliasing or value forwarding).
%
When the processor realizes it has mispredicted,
it \emph{rolls back} execution, erasing the programmer-visible effects
of the speculation.
However,
\emph{microarchitectural} state---such as the state of the data cache---is
still modified during speculative execution; these changes can be
leaked during speculation and can persist even after rollback.
As a result, the attacker can recover sensitive information from the
microarchitectural state, even if the sensitive information was only
speculatively accessed.

\Cref{fig:crypto_code_example_data_cache} gives an example of a
vulnerable function: An attacker can exploit branch misprediction
to leak arbitrary memory via the data cache.
The attacker first primes the branch to predict that the condition
\ccode{i < arrALen} is true by causing the code to repeatedly run with
appropriate (small) values of \ccode{i}.
Then, the attacker provides an out-of-bounds value for \ccode{i}.
The processor (mis)predicts that the condition is still true and
\emph{speculatively} loads out-of-bounds (potentially secret) data into \ccode{x};
subsequently, it uses the value \ccode{x} as part of the address of a memory
read operation.
This encodes the value of \ccode{x} into the data cache state---%
depending on the value of \ccode{x}, different cache lines will be
accessed and cached.
Once the processor resolves the misprediction, it rolls back execution,
but the data cache state persists.
The attacker can later interpret the data cache state in order to infer the
value of \ccode{x}.

\subsection{Breaking cryptography with Spectre}
\label{subsec:prelims-ct}

\begin{figure}
\begin{cblockindent}
if (i < arrALen) {  // mispredicted
  int x = arrA[i];  // x is oob value
  int y = arrB[x];  // leaked via address!
  // ...
\end{cblockindent}
\caption{
  Code snippet which an attacker can exploit using Spectre.
  If an attacker can control \ccode{i} and cause the processor to transiently
  enter the branch, the attacker can load an arbitrary value from memory into \ccode{x},
  which is then leaked via the following memory access.
}
\label{fig:crypto_code_example_data_cache}
\end{figure}

High-assurance cryptography has long relied on
\emph{constant-time programming}~\cite{barthe2014}
in order to create software which is secure from timing side-channel attacks.
%
Constant-time programming ensures that program execution does not depend on
secrets.
It does this via three rules of thumb~\cite{barthe2014,barthe2018}: control
flow (\eg conditional branches) should not depend on secrets, memory access
patterns (\eg offsets into arrays) should not be influenced by secrets, and secrets
should not be used as operands to variable-latency instructions (\eg
floating-point instructions or integer division on many processors).
These rules ensure that secrets remain safe from an attacker powerful enough to
perform cache attacks, exfiltrate data via branch predictor state, or
snoop data via port contention~\cite{smotherspectre}.

In the face of Spectre, constant-time programming is not sufficient.
The snippet in
\Cref{fig:crypto_code_example_data_cache} is indeed constant-time if
\ccode{arrA} contains only public data (and \ccode{i} and \ccode{arrALen}
are also public).
Yet, a Spectre attack can still abuse this code to leak secrets from
anywhere in memory.

Cache-based leaks are not the only way for an attacker to
learn cryptographic secrets:
In the following example, an attacker can again
(speculatively) leak out-of-bounds data, but this time the leak is via
control flow.
\begin{cblock}
if (i < arrALen) {
  int x = arrA[i];
  switch(x) {  // leak via branching!
    case 'A': /* ... */
    case 'B': /* ... */
    // ...
\end{cblock}
This code uses \ccode{x} as part of a branch condition (in a \ccode{switch}
statement).
Just as before, the attacker can speculatively read arbitrary memory into \ccode{x}.
They can then leak the value of \ccode{x} in several ways, including:
(1) Based on the different execution times of the various cases;
(2) through the data cache, based on differing (benign) memory accesses performed in the various cases;
(3) through the instruction cache \NEW{or micro-op cache~\cite{iseedeaduops}}, based on which instructions were (speculatively) accessed; 
or (4) through port contention~\cite{smotherspectre}, \NEW{branch predictor state~\cite{branchspec}}, or other microarchitectural resources that differ among the~branches.


\subsection{Breaking software isolation with Spectre}
\label{subsec:prelims-isolation}
Spectre attacks also break important guarantees in the domain of
\emph{software isolation}.
In this domain, a host application executes untrusted code and
wants to ensure that the untrusted code cannot access any of the host's data.
Common examples of software isolation include JavaScript or WebAssembly
runtimes, or even the Linux kernel, through eBPF~\cite{fleming2017ebpf}.
Spectre attacks can break the memory safety and isolation mechanisms commonly
used in these settings~\cite{splitspectre,swivel,venkman,jenkins2020ghostbusting}.

We demonstrate with a small example:
\begin{cblock}
int guest_func() {
  get_host_val(1);
  get_host_val(1);
  // ... repeat ...
  char c = get_host_val(99999);
  // ... leak c
}

char get_host_val(int idx) {
  if (idx < 100) { // check if within bounds
    return host_arr[idx];
  } else {
    return 0;
} }
\end{cblock}
Here, an attacker-supplied guest function \ccode{guest_func} calls
the host function \ccode{get_host_val} to get values from an array.
Although \ccode{get_host_val()} implements a bounds check, the attacker
can still speculatively access out-of-bounds data by mistraining the branch
predictor---breaking any isolation guarantees.
Once the attacker (speculatively) obtains an out-of-bounds value of their
choosing, they can leak the value (\eg via data cache, etc.)
and recover it after the speculative rollback.
In this setting, we need to ensure
that, \emph{even speculatively}, untrusted code cannot
break isolation.

\subsection{Security properties and execution semantics}
\label{subsec:prelims-formal}

Formally, we will
define safety from Spectre attacks as a security property
of a \emph{formal (operational) semantics}.
%
The semantics abstractly captures how a processor executes a program as a series of
state transitions.
%
%
The states, which we will write as $\sigma$, include any information the developer will
need to track for their analysis,
such as the current instruction or command and the contents
of memory and registers.
The developer then
defines an \emph{execution model}---a set of transition rules that specify how state
changes during execution.
For example, in a semantics for a low-level assembly, a rule for a \ccode{store}
instruction will update the resulting state's memory with a new value.

The rules in the execution model
determine how and when speculative effects happen.
For example, in a sequential semantics,
  a conditional branch will evaluate its condition
    then step to the appropriate branch.
    A semantics that models branch prediction will instead \emph{predict}
    the condition result and step to the predicted branch.
  We adapt notation from Guarnieri \etal~\cite{guarnieri2021contracts},
  writing \contract{seq}{} to represent the execution model for standard sequential execution.
  We notate other execution models similarly; for example, \contract{pht}{}
  models prediction for Spectre-PHT attacks---\ie conditional branch prediction.
  Other execution models are listed in \Cref{tab:props}.

Next, to precisely specify the attacker model, the developer
must define which
\emph{leakage observations}---information
produced during an execution step---are visible to an attacker. For example,
they may decide that rules with memory accesses leak the addresses being accessed.
The set of leakage observations in a semantics' rules is its
\emph{leakage model}.
We again borrow notation from Guarnieri \etal~\cite{guarnieri2021contracts},
which defines the leakage models \contract{}{ct} and \contract{}{arch}.
The \contract{}{ct} model exposes leakage observations relevant to constant-time security:
The sequence of control flow (the \emph{execution trace}) and the sequence
of addresses accessed in memory (the \emph{memory trace}).%
\footnote{Like Guarnieri \etal~\cite{guarnieri2021contracts}, we omit variable-latency
instructions from our formal model for simplicity.}
The \contract{}{arch} model, on the other hand, exposes
all values loaded from memory in addition to the addresses themselves (or equivalently,
it exposes the trace of register values)~\cite{guarnieri2021contracts}.
\NEW{
Under this model, an attacker is allowed to observe all architectural computation;
for a value to remain unobserved, it cannot be accessed at all over the course of execution,
adversarial or otherwise.
}
Since the leakage observations in \contract{}{arch} are a strict superset of
those in \contract{}{ct}, we say that \contract{}{arch} is \emph{stronger}
than \contract{}{ct} (\ie it models a more powerful attacker).
%
%
\NEW{
These properties make \contract{}{arch} most useful for software isolation,
as any out-of-bounds accesses will immediately show up in an \contract{}{arch} leakage trace.
}
%

\NEW{
Surprisingly, the \contract{}{ct} and \contract{}{arch} leakage models
generalize well to speculative execution---for example, if we want to construct a semantics for Spectre-PHT
attacks, we need only modify a sequential constant-time semantics to account
for branch misprediction.
Indeed, the execution model and leakage model of a semantics are orthogonal;
we call the combination of the two the \emph{contract} provided by the semantics---%
a sequential constant-time semantics has the contract \contract{seq}{ct},
while our hypothetical Spectre-PHT semantics would provide the contract \contract{pht}{ct}.}
Formally, the contract governs the attacker-visible information produced when executing a program:
Given a program $p$, a semantics with contract \contract{$\alpha$}{$\ell$}, and an initial state $\sigma$, we write
  $\contractP{$\alpha$}{$\ell$}(\sigma)$ for the sequence (or \emph{trace}) of leakage observations
  the semantics produces
  when executing $p$.


  After determining a proper contract,
the developer must finally define the \emph{policy} that their security property enforces:
Precisely which data can and cannot
be leaked to the attacker.
Formally, a policy $\pi$ is defined in terms of an equivalence relation $\simeq_\pi$ over states,
where $\sigma_1 \simeq_\pi \sigma_2$
iff $\sigma_1$ and $\sigma_2$ agree on all values that are public (but may differ on sensitive values).

\NEW{
Armed with these definitions, we can state security as a
\emph{non-interference property}:
A program satisfies \emph{non-interference} if, for any two $\pi$-equivalent initial states for a program $p$,
an attacker cannot distinguish the two resulting leakage traces when executing $p$.}
%
%
A developer has several choices when crafting a suitable semantics and
security policy; these choices greatly influence how easy or difficult it
is to detect or mitigate Spectre vulnerabilities.
We cover these choices in detail in \Cref{sec:semantics}:
\Cref{subsec:leakage,subsec:props} discuss choices in leakage models
\contract{}{$\ell$} and security policies $\pi$.
\Cref{subsec:predictors,subsec:nondeterminism} discuss tradeoffs for different
execution models \contract{$\alpha$}{} and the transition rules in a semantics.
In \Cref{subsec:low-vs-high}, we discuss how the input language of the semantics
affects analysis; and finally, in \Cref{sec:expressivity}, we discuss
which microarchitectural features
to include in formal models.
%

\section{Choices in semantics}
\label{sec:semantics}

The foundation of a well-designed Spectre analysis tool is a
carefully constructed formal semantics.
Developers face a wide variety of choices when designing their
semantics---choices which heavily depend on the attacker model (and thus the
intended application area) as well as specifics about the tool they want to
develop.
Cryptographic code requires different security properties, and therefore
different semantics and tools, than in-process isolation.
Many of these choices also look different for \emph{detection} tools, focused
only on finding Spectre vulnerabilities, vs.\@ \emph{mitigation} tools, which
transform programs to be secure.
In this section, we describe the important choices about semantics that
developers face, and explain those choices' consequences for Spectre analysis
tools and for their associated security guarantees.
We also point out a number of open problems to guide future work in this
area.

\myparax{What makes a practical semantics?}
%
%
A practical semantics should make an appropriate tradeoff
between \emph{detail} and \emph{abstraction}: 
It should be detailed enough to capture the microarchitectural behaviors
which we're interested in, but it should also be abstract enough that it
applies to all (reasonable) hardware.
For example, we do not want the security of our code to
be dependent on a specific cache replacement policy or branch predictor
implementation.

In the non-speculative world, formalisms for constant-time have been
successful:
The principles of constant-time programming
(no secrets for branches, no secrets for addresses)
create secure code without introducing processor-specific abstractions.
%
Speculative semantics should follow this trend, producing
portable tools which can defend against powerful attackers on today's (and
tomorrow's) microarchitectures.

\subsection{Leakage models}
\label{subsec:leakage}

Any semantics intended to model side-channel attacks needs to precisely
define its attacker model.
An important part of the attacker model for a semantics is the \emph{leakage
model}---that is, what information does the attacker get to observe?
Leakage models intended to support sound mitigation schemes should be
\emph{strong}---modeling a powerful attacker---and \emph{hardware-agnostic},
so that security guarantees are portable.
That said, the best choice for a leakage model depends in large part on the
intended application domain.

\begin{figure*}
  \footnotesize
  \centering

  \newcommand{\fnmark}[1]{{\small\textsuperscript{#1}}}

  \newcommand{\toolname}[1]{\scriptsize (#1)}
  \newcommand{\yes}{\checkmark\xspace}
  \newcommand{\no}{{\bfseries $\times$\xspace}}
  \newcommand{\kindof}{{\bfseries $\backsim$\xspace}}
  \newcommand{\forward}{$\rightarrow$}
  \newcommand{\key}[2]{\emph{#1}~-- #2}

  \newcommand{\q}{+}
  \newcommand{\taint}{Taint}
  \newcommand{\cache}{Cache}
  \newcommand{\relCache}{\cache\q}
  \newcommand{\manual}{Manual}
  \newcommand{\safety}{Safety}
  \newcommand{\flow}{Flow}
  \newcommand{\SC}{SelfC}
  \newcommand{\relSC}{\SC\q}
  \newcommand{\compiler}{Struct}
  \newcommand{\fuzz}{Fuzz}
  \newcommand{\model}{Model}

  \begin{tabular}{lll@{ }lll>{\centering}b{1em}>{\centering}b{1em}>{\centering}b{1em}>{\centering}b{1em}b{2.5em}l}
    Semantics or tool name & Level & \multicolumn{2}{l}{Leakage} & Variants & Nondet. & Fence & OOO & Win. & Hij. & Tool & Impl. \\
    \toprule
    Cauligi \etal~\cite{cauligi2020foundations} \toolname{Pitchfork}
    & Low & \contract{}{ct} & P,B,M & P,B,R,S & Directives
    & \yes & \yes & \yes & \yes & Det* & \taint \\
    \cmidrule(lr){1-12}
    Cheang \etal~\cite{cheang2019}
    & Low & \contract{}{arch} & P,M,S,R & P & Oracle
    & \yes & \no & \yes & \no & Det/Mit & \relSC \\
    \cmidrule(lr){1-12}
    Daniel \etal~\cite{hauntedrel} \toolname{Binsec/Haunted}
    & Low & \contract{}{ct} & P,M & P,S & Mispredict
    & \no & \no & \yes & \no & Det & \SC \\
    \cmidrule(lr){1-12}
    Guanciale \etal~\cite{guanciale2020inspectre} \toolname{InSpectre}
    & Low & \contract{}{ct} & P,M & P,B,R,S & ---
    & \yes & \yes & \no & \yes & --- & --- \\
    \cmidrule(lr){1-12}
    Guarnieri \etal~\cite{guarnieri2020spectector} \toolname{Spectector}
    & Low & \contract{}{ct} & P,B,M & P & Oracle
    & \yes & \no & \yes & \forward & Det & \relSC \\
    \cmidrule(lr){1-12}
    Guarnieri \etal~\cite{guarnieri2021contracts}
    & Low & \multicolumn{2}{l}{(parametrized)} & P\fnmark{1} & Oracle
    & \yes & \yes & \yes & \no & Det & \relSC \\
    \cmidrule(lr){1-12}
    Mcilroy \etal~\cite{mcilroy2019}
    & Low & \contract{}{cache} & T & P\fnmark{2} & Oracle
    & \kindof & \no & \yes & \forward & Mit* & \manual \\
    \cmidrule(lr){1-12}
    Barthe \etal~\cite{jasmin} \toolname{Jasmin}
    & Medium & \contract{}{ct} & P,B,M & P,S & Directives
    & \yes & \no & \no & \no & Det & \safety \\
    \cmidrule(lr){1-12}
    Patrignani and Guarnieri~\cite{patrignani2020}
    & Medium & \contract{}{ct} & P,B,M,L\fnmark{3} & P\fnmark{1} & Mispredict
    & \yes & \no & \yes & \no & --- & --- \\
    \cmidrule(lr){1-12}
    Vassena \etal~\cite{vassena2021} \toolname{Blade}
    & Medium & \contract{}{ct} & B,M & P & Directives
    & \yes & \yes & \no & \no & Mit & \flow \\ 
    \cmidrule(lr){1-12}
    Colvin and Winter~\cite{colvin2019}
    & High & \contract{}{mem} & M & P & Weak-mem
    & \yes & \yes & \no & \no & Val & \model \\
    \cmidrule(lr){1-12}
    Disselkoen \etal~\cite{disselkoen2019}
    & High & \contract{}{mem} & M & P & Weak-mem
    & \yes & \yes & \no & \no & --- & --- \\
    \cmidrule(lr){1-12}
    P. de Le\'on and Kinder~\cite{cats-vs-spectre} \toolname{Kaibyo}
    & High & \contract{}{mem} & M & P,S & Weak-mem
    & \yes & \yes & \yes & \no & Det & \model \\
    \midrule
    AISE~\cite{wu-wang}
    & --- & \contract{}{cache} & C & P & Mispredict
    & \no & \no & \yes & \no & Det & \relCache \\
    \cmidrule(lr){1-12}
    \NEW{ASTCVW~\cite{spectypeconfusion} }
    & --- & \contract{}{arch} & L & P\fnmark{4} & ---
    & \no & \no & \no & \no & Det & \taint \\
    \cmidrule(lr){1-12}
    ELFbac~\cite{jenkins2020ghostbusting}
    & --- & \contract{}{arch} & L & P & ---
    & \no\fnmark{5} & \no & \no & \yes & Mit & \compiler \\
    \cmidrule(lr){1-12}
    \multirow{2.5}{*}{KLEESpectre~\cite{kleespectre}} \hfill\toolname{w/ cache}
      & --- & \contract{}{cache} & C & P & Mispredict
      & \yes & \no & \yes & \no & Det & \cache \\
    \cmidrule(r){2-12}
    \hphantom{KLEESpectre~\cite{kleespectre}} \hfill\toolname{w/o cache}
      & --- & \contract{}{mem} & M & P & Mispredict
      & \yes & \no & \yes & \no & Det & \taint \\
    \cmidrule(lr){1-12}
    \multirow{2.5}{*}{oo7~\cite{oo7}} \hfill\toolname{v1 pattern}
      & --- & \contract{}{mem} & M & P & ---
      & \kindof & \no & \yes & \no & Det/Mit & \flow \\
    \cmidrule(r){2-12}
    \hphantom{oo7~\cite{oo7}} \hfill\toolname{``weak'' and v1.1 patterns}
      & --- & \contract{}{arch} & L & P & ---
      & \kindof & \no & \yes & \kindof & Det/Mit & \flow \\
    \cmidrule(lr){1-12}
    \NEW{Specfuscator~\cite{specfuscator} }
    & --- & ---\fnmark{6} & --- & P,B,R & ---
    & \no\fnmark{5} & \no & \no & \yes & Mit & \compiler \\
    \cmidrule(lr){1-12}
    SpecFuzz~\cite{specfuzz}
    & --- & \contract{}{arch} & L & P & Mispredict
    & --- & --- & --- & \yes & Det & \fuzz \\
    \cmidrule(lr){1-12}
    \NEW{SpecTaint~\cite{spectaint} }
    & --- & \contract{}{mem}\fnmark{7} & M & P & Mispredict
    & \yes & \no & \yes & \kindof & Det & \taint \\
    \cmidrule(lr){1-12}
    SpecuSym~\cite{specusym}
    & --- & \contract{}{cache} & C & P & Mispredict
    & \no & \no & \yes & \no & Det & \relSC \\
    \cmidrule(lr){1-12}
    \multirow{2.5}{*}{Swivel~\cite{swivel}} \hfill\toolname{poisoning protection}
      & --- & \contract{}{mem} & M & P,B,R & ---
      & \kindof\fnmark{8} & \no & \no & \yes & Mit & \compiler \\
    \cmidrule(r){2-12}
    \hphantom{Swivel~\cite{swivel}} \hfill\toolname{breakout protection}
      & --- & \contract{}{arch} & L & P,B,R & ---
      & \kindof\fnmark{8} & \no & \no & \yes & Mit & \compiler \\
    \cmidrule(lr){1-12}
    Venkman~\cite{venkman}
    & --- & \contract{}{arch} & L & P,B,R & ---
    & \kindof & \no & \no & \yes & Mit & \compiler \\
    \bottomrule
  \end{tabular}

  {\scriptsize
    \vspace{1em}
    \begin{tabular}{l@{\hspace{0.25em}}l l@{\hspace{0.25em}} l@{\hspace{0.75em}} l@{\hspace{0.25em}}l  l@{\hspace{0.25em}}l }
      \multicolumn{2}{l}{Level -- How abstract is the semantics? (\Cref{subsec:low-vs-high})}
      & \multicolumn{4}{l}{Leakage -- What can the attacker observe? (\Cref{subsec:leakage})}
      & \multicolumn{2}{l}{Variants (\Cref{subsec:predictors})}
      \\\cmidrule(l{0em}r{0em}){1-8}
      Low & Assembly-style, with branch instructions
      & P &-- Path / instructions executed       & L &-- Values loaded from memory          & P &-- Spectre-PHT                     \\
      Medium & Structured control flow such as if-then-else
      & B &-- Speculation rollbacks              & R &-- Values in registers                & B &-- Spectre-BTB                     \\
      High & In the style of weak memory models
      & M &-- Addresses of memory operations     & S &-- Branch predictor state             & R &-- Spectre-RSB                     \\
      --- & The work has no associated formal semantics
      & C &-- Cache lines / cache state         & T &-- Step counter / timer               & S &-- Spectre-STL                     \\
    \end{tabular} \\
    \vspace{.5em}
    \begin{tabular}{l@{\hspace{0.5em}} m{0.5\textwidth} @{\hspace{1em}} l@{\hspace{0.5em}} m{0.35\textwidth}}
      \multicolumn{2}{l}{Fence -- Does it reason about speculation fences?} &
      \multicolumn{2}{l}{\hspace{-1em}Hijack -- Can it model or mitigate speculative hijack?}
      \\\cmidrule(l{0em}r{0em}){1-4}
      \yes & Fully reasons about fences in the target/input code &
      \yes & Models/mitigates speculative hijack attacks \\
      \multirow{2}{*}{\kindof} & The mitigation tool inserts fences, but the analysis does not reason about fences &
      \forward & Models/mitigates forward-edge (ijmp) hijack only \\
       & in the target/input code (and thus cannot verify the mitigated code as secure) &
      \kindof & Models/mitigates hijack only via speculative stores \\
      \no & Does not reason about, or insert, fences &
      \no & Does not model/mitigate speculative hijack attacks \\
    \end{tabular} \\
    \vspace{.5em}
    \begin{tabular}{l}
      Nondet. -- How is nondeterminism handled? (\Cref{subsec:nondeterminism}) \\
      OOO -- Models out-of-order execution? (\Cref{subsec:ooo}) \\
      Win. -- Can reason about speculation windows? (\Cref{subsec:predictors}) \\
    \end{tabular} \\
    \vspace{.5em}
    \begin{tabular}{l@{\hspace{0.5em}} m{0.38\textwidth} l@{\hspace{0.8em}} l@{\hspace{2.5em}} l@{\hspace{0.8em}}l}
      \multicolumn{2}{l}{Tool -- Does the paper include a tool?} &
      \multicolumn{4}{l}{Implementation -- How does the tool detect or mitigate vulnerabilities? (\Cref{subsec:nondeterminism})} \\
      \cmidrule(l{0em}r{0em}){1-6}
      Det & Tool detects insecure programs or verifies secure programs &
      \taint & Taint tracking (abstract execution) &
      \manual & Manual effort \\
      Mit & Tool modifies programs to ensure they are secure &
      \safety & Memory safety (abstract execution) &
      \fuzz & Fuzzing \\
      \multirow{2}{*}{Val} & Tool is only used to validate the semantics, does not &
      \SC & Self composition (abstract execution) &
      \flow & Data flow analysis \\
      & automatically perform any security analysis &
      \cache & Cache must-hit analysis (abstract execution) &
      \compiler & Structured compilation \\
      \multicolumn{1}{r}{---} & Does not include a tool &
      \model & Model checking over the whole program \\
      \multicolumn{1}{c}{\hspace{0.2em}*} & Tool's connection to the semantics is incomplete or unclear (e.g., tool does not implement the full semantics) & 
      \multicolumn{4}{c}{\q \hspace{1em} Includes additional work or
            constraints to remove sequential trace (\Cref{subsec:props})} \\
    \end{tabular}
  }
  \caption{
    Comparison of various semantics and tools.
    Semantics are sorted by \emph{Level}, then alphabetically; works without semantics are ordered last.
    \fnmark{1}Extension to other variants is discussed, but not performed.
    \fnmark{2}Semantics includes indirect jumps and rules to update the indirect branch predictor state, but cannot mispredict indirect jump targets.
    \fnmark{3}``Weak'' variants of semantics leak loaded values during non-speculative execution.
    \NEW{
    \fnmark{4}Detects only ``speculative type confusion vulnerabilities'', a specific subset of Spectre-PHT.
    }
    \fnmark{5}Mitigates Spectre-PHT without inserting fences.
    \NEW{
    \fnmark{6}Defends by effectively preventing speculation, so leakage model is irrelevant.
    \fnmark{7}Effectively {\scriptsize\contract{}{mem}} for loads, but detects any speculative store to an attacker-controlled address, which is more similar to {\scriptsize\contract{}{arch}} for stores.
    }
    \fnmark{8}Swivel operates on WebAssembly, which does not have fences. However, Swivel can insert fences in its assembly backend.
  }
  \label{tab:mechanics}
\end{figure*}

\mypara{Leakage models for cryptography}
As we saw in \Cref{subsec:prelims-ct}, high-assurance cryptography
implementations have long relied on the constant-time programming model;
thus, semantics intended for cryptographic programs naturally choose the
\contract{}{ct} leakage model.
Like the constant-time programming model in the non-speculative world, the
\contract{}{ct} leakage model is strong and hardware-agnostic, making it a
solid foundation for security guarantees.
The \contract{}{ct} leakage model is a popular choice among existing
formalizations:
As we highlight in \Cref{tab:mechanics}, over half of the formal
semantics for Spectre use the \contract{}{ct} leakage model (or an
equivalent)~\cite{jasmin,cauligi2020foundations,hauntedrel,guanciale2020inspectre,guarnieri2020spectector,patrignani2020,vassena2021}.
Guarnieri \etal~\cite{guarnieri2021contracts}
leave the leakage model abstract, allowing the semantics to be used with
several different leakage models, including~\contract{}{ct}.


\mypara{Leakage models for isolation}
Sections~\ref{subsec:prelims-isolation} and~\ref{subsec:prelims-formal}
describe the \contract{}{arch} leakage model, which is a better fit for modeling speculative
isolation, \NEW{\eg for a WebAssembly runtime executing untrusted
code~\cite{swivel} or a kernel defending against memory region probing~\cite{specprobing}.
%
Under \contract{}{arch}, \emph{all} values in the program are observable---%
this is what lets it easily model properties for software isolation:
If we define a policy $\pi$ where all values and memory regions outside the isolation boundary
are secret, then software isolation security (or speculative memory safety) is simply non-interference with respect
to $\contract{}{arch}$ (and this $\pi$).
}
%

The \contract{}{arch} leakage model appears less frequently than \contract{}{ct}
in formal models:
Only two of the semantics in \Cref{tab:mechanics}
(\cite{cheang2019,guarnieri2021contracts}) use the \contract{}{arch} leakage
model.
%
%
%
On the other hand,
Spectre sandbox isolation frameworks such as Swivel~\cite{swivel},
Venkman~\cite{venkman}, and ELFbac~\cite{jenkins2020ghostbusting}
implicitly use the \contract{}{arch} model, as do
the detection tools
SpecFuzz~\cite{specfuzz}, \NEW{ASTCVW~\cite{spectypeconfusion}},
\NEW{SpecTaint~\cite{spectaint}}, and the ``weak'' and ``v1.1'' modes of oo7~\cite{oo7}.
\NEW{
The three isolation frameworks all explicitly prevent memory reads or writes
to any locations outside of the isolation boundary---\ie enforcing non-interference under \contract{}{arch}.
The detection tools,
meanwhile,
look for gadgets that can speculatively access \emph{arbitrary}
(or attacker-controlled) memory locations---\ie breaking speculative memory
safety.
}
%
%
Unfortunately, these tools are not formalized, so their leakage
models are not made explicit (nor clear).

\mypara{Weaker leakage models}
The remaining semantics and tools in \Cref{tab:mechanics} consider only the memory
trace of a program, but not its execution trace.
The \contract{}{mem} leakage model, like \contract{}{ct}, allows an attacker
to observe the sequence of memory accesses during the execution of the program;
the \contract{}{cache} leakage model instead only tracks (an abstraction of) cache state.
The attacker in this model can only observe cached addresses
at the granularity of cache lines.
%
%
A few tools have even weaker leakage models---for instance, oo7
only emits leakages that can be influenced by malicious input
(see \Cref{marker:in-out-place}) and KLEESpectre (with cache modeling
enabled) only allows the attacker to observe the final state of the cache
upon termination. 

All of these models, including \contract{}{mem} and \contract{}{cache}, are
weaker than \contract{}{ct}---they model less powerful attackers who cannot
observe control flow.
As a result, they miss attacks which leak via the instruction cache or which
otherwise exploit timing differences in the execution of the program.
They even miss some attacks that exploit the data cache:
If a sensitive value influences a branch, an attacker could infer the
sensitive value through the data cache based on differing (benign) memory
access patterns on the two sides of the branch, even if no sensitive value
directly influences a memory address.
For instance, in the following code, even though \ccode{cond}
is not used to calculate the memory address, an attacker can infer the value of
\ccode{cond} based on whether \ccode{arr[a]} gets cached or not:
\begin{cblock}
if (cond)
  b = arr[a];
else
  b = 0;
\end{cblock}
Because the \contract{}{mem} and \contract{}{cache} leakage models miss these
attacks, they cannot provide the strong
guarantees necessary for secure cryptography or software isolation.
Tools which want to provide sound verification or mitigation should instead choose a
strong leakage model appropriate for their application domain, such as
\contract{}{ct} or \contract{}{arch}.

That said, weaker leakage models are still useful in certain settings:
Tools which are interested in only certain vulnerability classes can use
these weaker models to reduce the number of false positives in their analysis
or reduce the complexity of their mitigation.
Even though these models may miss some Spectre attacks,
detection tools can still use the
\contract{}{cache} or \contract{}{mem} models to find Spectre vulnerabilities
in real codebases.
Using a leakage model which ignores control flow leakage may help the
detection tool scale to larger codebases.
%

Some tools~\cite{specusym,kleespectre}
also provide the ability to reason about what attacks are
possible with particular cache configurations---\eg with a particular
associativity, cache size, or line size.
This is a valuable capability for a detection tool:
It helps an attacker zero in on vulnerabilities which are more easily
exploitable on a particular target machine.
However, security guarantees based on this kind of analysis are not
portable, as executing a program on a different machine with a different cache
model invalidates the security analysis.
Tools that instead want to make guarantees for all possible architectures,
such as verifiers or compilers, will
need more conservative leakage models---models
that assume the entire memory trace (and execution
trace) is always leaked.

%


\mypara{Open problems: Leakage models for weak-memory-style semantics}
We have described leakage models only in terms of observations of execution
traces; this is a natural way to define leakage for \emph{operational
semantics}, where execution is modeled simply as a set of program traces.
However, the weak-memory-style speculative semantics proposed by Colvin and Winter~\cite{colvin2019},
Disselkoen \etal~\cite{disselkoen2019}, and Ponce de Le\'on and Kinder~\cite{cats-vs-spectre}
have a more structured view of program execution (for instance, using dependency analysis or pomsets~\cite{pomsets}).
These semantics define leakage equivalent to the
\contract{}{mem} leakage model; it remains an open problem to explore how to
define \contract{}{ct} or \contract{}{arch} leakage in this more structured
execution model---in particular, what it means for such a semantics to allow
an attacker to observe control-flow.

\mypara{Open problems: Leakage models for language-based isolation}
As with most work on Spectre foundations, we focus on cryptography and
software-based isolation. Spectre, though, can be used to break most other software
abstractions as well---from module systems~\cite{wasm} and object capabilities~\cite{ocap}
to language-based isolation techniques like information flow control~\cite{sabelfeld2003ifc}.
How do we adopt these abstractions in the presence of speculative execution?
What formal security property should we prove?
And what leakage model should be used?

\subsection{Non-interference and policies}
\label{subsec:props}

After the leakage model, we must determine what \emph{secrecy policy} we consider
for our attacker model---\ie which values can and cannot be leaked.
Domains such as cryptography and isolation already have defined policies for
sequential security properties: For cryptography, memory that contains secret data
(\eg encryption keys) is considered sensitive; isolation simply declares that all memory
outside the program's assigned sandbox region should not be leaked.
%

The straightforward extension of sequential non-interference
to speculative execution is to enforce the same leakage model (\eg \contract{}{ct})
with the same security policy---no secrets should
be leaked whether in normal or speculative execution.
%
We refer to this simple extension as a \emph{direct}
non-interference property, or \emph{direct NI}.

\begin{definition}[Direct non-interference]
  \label{def:direct-ni}
  Program $\p$ satisfies \emph{direct non-interference} with respect to a given contract \contract{}{}
  and policy $\pi$ if, for all pairs of $\pi$-equivalent initial states $\sigma$ and $\sigma'$,
  executing $\p$ with each initial state produces the same trace.
  That is,
  $\p \satisfies \nipi{\pi}{\contract{}{}}$ is defined as
  \[ \forall \sigma,\sigma' : \sigma \simeq_\pi \sigma' \implies \contractP{}{}(\sigma) = \contractP{}{}(\sigma'). \]
\end{definition}
\noindent
We elide writing $\pi$ for brevity---\eg $\ni{\contract{pht}{ct}}$ expresses constant-time security
under Spectre-PHT semantics.

Alternatively, we may instead want to assert that the speculative trace
of a program
has no \emph{new} sensitive leaks as compared to its sequential trace.
%
This is a useful property for compilers and mitigation tools that may
not know the secrecy policy of an input program, but want to ensure
the resulting program does not leak any additional information.
We term this a \emph{relative} non-interference property, or \emph{relative NI}; a
program that satisfies relative NI is no less secure than its sequential
execution.

%
\begin{definition}[Relative non-interference]
  \label{def:relative-ni}
  Program $\p$ satisfies \emph{relative non-interference} from contract \contract{seq}{a} to \contract{$\beta$}{b}
  and with policy $\pi$ if: For all pairs of $\pi$-equivalent initial states $\sigma$ and $\sigma'$,
  if executing $\p$ under \contract{seq}{a} produces equal traces, then executing $\p$ under \contract{$\beta$}{b} produces equal traces.
  That is,
  $\p \satisfies \nipi{\pi}{\contract{seq}{a} \implies \contract{$\beta$}{b}}$ is defined as
  \begin{align*}
    \forall \sigma,\sigma' : \sigma \simeq_\pi \sigma' \land \contractP{seq}{a}(\sigma) = \contractP{seq}{a}(\sigma') \\ \Longrightarrow \contractP{$\beta$}{b}(\sigma) = \contractP{$\beta$}{b}(\sigma').
  \end{align*}
\end{definition}
\noindent
\NEW{
For non-terminating programs, we can compare finite prefixes of \contractP{$\beta$}{} against their sequential
projections to \contractP{seq}{}---since speculative execution must preserve sequential semantics,
there will always be a valid sequential projection.
}
As before, we may elide $\pi$ for brevity.

Interestingly,
any relative non-interference property $NI(\pi, \contract{seq}{a} \implies \contract{$\beta$}{b})$
for a program $p$
can be expressed equivalently as a direct property $NI(\pi', \contract{$\beta$}{b})$, where
$\pi' = \pi \setminus canLeak(p, \contract{seq}{a})$. That is, we treat anything that could possibly
leak under contract \contract{seq}{a} as public. Relative NI
is thus a (semantically) weaker property than direct NI, as it implicitly
declassifies anything that might leak during sequential execution.


However, relative NI is still a stronger property than a conventional
implication.
For example, the property $\ni{\contract{seq}{ct}} \implies \ni{\contract{pht}{ct}}$
makes no guarantees at all about a program that is not sequentially constant-time.
Conversely, the relative NI property $\ni{\contract{seq}{ct} \implies \contract{pht}{ct}}$
guarantees that even if a program is not sequentially constant-time,
the sensitive information an attacker can learn during the program's speculative execution
is limited to what it already might leak sequentially.

In \Cref{tab:props}, we classify security properties of different works
by which direct or relative NI properties they verify or enforce.
We find that tools focused on verifying cryptography or memory isolation
verify direct NI properties, whereas frameworks concerned with
compilation or inserting Spectre mitigations for general programs
tend towards relative NI.

\begin{figure*}
  \small
  \centering

  \newcommand{\fnmark}[1]{{\small\textsuperscript{#1}}}

  \newcommand{\yes}{\checkmark\xspace}
  \newcommand{\no}{{\bfseries $\times$\xspace}}
  \newcommand{\kindof}{{\bfseries $\backsim$\xspace}}

  \newcommand{\hyper}{hyper}
  \newcommand{\taint}{taint}
  \newcommand{\hypertaint}{hyper, taint}

  \begin{tabular}{lll}
    Property or tool name & Non-interference prop. & Precision \\\toprule
    Mcilroy \etal~\cite{mcilroy2019}
      & $\approx$\ni{\contract{pht}{ct}}
      & hyper
      \\ \cmidrule(lr){1-3}
    \multirow{2}{*}{oo7~\cite{oo7}} $\Phi_{spectre}$
      & $\approx$\ni{\contract{pht}{mem}}
      & \multirow{2}{*}{\taint\fnmark{1}}
      \\
    \hphantom{oo7~\cite{oo7}} $\Phi_{spectre}^{weak}$, $\Phi_{spectre}^{v1.1}$
      & $\approx$\ni{\contract{pht}{arch}}
      &
      \\\midrule
    \multirow{2}{*}{Cache analysis} \cite{specusym,wu-wang}
      & \multirow{2}{*}{\ni{\contract{pht}{cache}}}
      & \hyper
      \\
    \hphantom{Cache analysis} \cite{kleespectre}
      & 
      & \taint
      \\ \cmidrule(lr){1-3}
    \multirow{2}{*}{Weak memory modeling} \cite{colvin2019,disselkoen2019}
      & \ni{\contract{pht}{mem}}
      & \multirow{2}{*}{hyper} \\
    \hphantom{Weak memory modeling} \cite{cats-vs-spectre}
      & \ni{\contract{pht-stl}{mem}}
      & 
      \\ \cmidrule(lr){1-3}
    \hphantom{Speculative constant-time (SCT)\fnmark{2}} \cite{vassena2021}
      & \ni{\contract{pht}{ct}}
      & \taint
      \\\cmidrule(r){2-3}
    \multirow{1}{*}{Speculative constant-time (SCT)\fnmark{2}} \cite{jasmin,hauntedrel}
      & \ni{\contract{pht-stl}{ct}}
      & \hyper
      \\\cmidrule(r){2-3}
    \hphantom{Speculative constant-time (SCT)\fnmark{2}} \cite{cauligi2020foundations}
      & \ni{\contract{pbrs}{ct}}\fnmark{3}
      & \hypertaint
      \\\midrule
    Speculative non-interference (SNI) \cite{guarnieri2020spectector,guarnieri2021contracts}
      & \ni{\contract{seq}{ct} \implies \contract{pht}{---}}\fnmark{4}
      & \hyper
      \\\cmidrule(lr){1-3}
    Robust speculative non-interference (RSNI)~\cite{patrignani2020}
      & \multirow{2}{*}{\ni{\contract{seq}{ct} \implies \contract{pht}{ct}}}
      & \hyper
      \\
    Robust speculative safety (RSS) \cite{patrignani2020}
      & 
      & \taint
      \\\cmidrule(lr){1-3}
    Conditional noninterference~\cite{guanciale2020inspectre}
      & \ni{\contract{seq}{ct} \implies \contract{pbrs}{ct}}
      & \hyper
      \\\midrule
    Weak speculative non-interference (wSNI) \cite{guarnieri2021contracts}
      & \ni{\contract{seq}{arch} \implies \contract{pht}{---}}\fnmark{4,5}
      & \hyper
      \\ \cmidrule(lr){1-3}
    Weak robust speculative non-interference (RSNI$^-$) \cite{patrignani2020}
      & \multirow{3}{*}{\ni{\contract{seq}{arch} \implies \contract{pht}{ct}}}
      & \hyper
      \\
    Trace property-dependent observational determinism (TPOD) \cite{cheang2019}
      & 
      & \hyper
      \\
    Weak robust speculative safety (RSS$^-$) \cite{patrignani2020}
      & 
      & \taint
      \\\bottomrule
  \end{tabular}

  {\footnotesize
    \vspace{1.5em}
    \begin{tabular}{l@{\hspace{0.8em}}lll@{\hspace{0.8em}}l}
      \multicolumn{2}{l}{Execution models (\Cref{subsec:predictors})} &
      \hspace{0.25em} &
        \multicolumn{2}{l}{Precision of the defined security property} 
    \\\cmidrule{1-5}
      \contract{seq}{} & Sequential execution  &
        & \hyper & Non-interference hyperproperty, requires two $\pi$-equivalent executions \\
      \contract{pht}{} & Captures Spectre-PHT  &
        & \taint & Sound approximation using taint tracking, requires only one execution \\
      \contract{pht-stl}{} & Captures Spectre-PHT/-STL \\
      \contract{pbrs}{} & Captures Spectre-PHT/-BTB/-RSB/-STL \\
    \end{tabular}
    \vspace{1em}
  }

  \caption{
    Speculative security properties in prior works and their equivalent non-interference statements.
    We write $\approx$\ni{\cdots} for unsound approximations of non-interference properties.
    %
    \fnmark{1}\cite{oo7} tracks taint of \emph{attacker influence} rather than value sensitivity.
    \fnmark{2}These works all derive their property from the definition given in \cite{cauligi2020foundations} and share the same property name despite differences in execution mode.
    \fnmark{3}The analysis tool of \cite{cauligi2020foundations}, Pitchfork, only verifies the weaker property \ni{\contract{pht-stl}{ct}}.
    \fnmark{4}The definitions of SNI and wSNI are parameterized over the target leakage model.
    \fnmark{5}The definition of wSNI in~\cite{guarnieri2021contracts} does not require that the initial states be $\pi$-equivalent.
  }
  \label{tab:props}
\end{figure*}

%
%


\mypara{Verifying programs}
%
%
Direct NI unconditionally guarantees that sensitive data is not leaked,
whether executing sequentially or speculatively.
%
%
This makes it ideal for domains
that already have clear policies about what data is sensitive,
such as cryptography (\eg secret keys) or software isolation (\eg memory outside the sandbox).
Indeed, tools that target cryptographic
applications~(\cite{cauligi2020foundations,vassena2021,hauntedrel,jasmin}) all
verify that programs satisfy the direct \emph{speculative constant-time} (SCT)
property.

%
%

Additionally, we find that current tools that verify relative
NI~\cite{guarnieri2020spectector,cheang2019} are indeed capable of verifying
direct NI, but intentionally
add constraints to their respective checkers to ``remove'' sequential leaks
from their speculative traces. Although this is just as precise,
%
it is an open problem whether tools can verify relative NI for programs
without relying on a direct NI analysis.

\mypara{Verifying compilers}
On the other hand,
compilers and mitigation tools are better suited to verify or
enforce relative NI properties:
%
%
%
The compiler guarantees that its output program contains no
new leakages as compared to its input program.
This way, developers can reason about their
programs assuming a sequential model, and the compiler will mitigate any
speculative effects.
For instance, if a program $p$ is already sequentially constant-time \ni{\contract{seq}{ct}},
then a compiler that enforces
\ni{\contract{seq}{ct} \implies \contract{pht}{ct}} will compile $p$ to a program
that is \emph{speculatively} constant-time \ni{\contract{pht}{ct}}.
Similarly, if a program is properly sandboxed under sequential execution \ni{\contract{seq}{arch}} and is compiled
with a compiler that introduces no new $\mathit{arch}$ leakage, the resulting program will
remain sandboxed even under speculative execution~\cite{guarnieri2021contracts}.

%

\NEW{
\label{marker:robustness}
Similarly, Patrignani and Guarnieri~\cite{patrignani2020} explore whether compilers
preserve \emph{robust} non-interference properties.
A security property is \emph{robust} if a program remains secure even when linked
against adversarial code (\ie if the program is called with arbitrary or
adversarial inputs).
A compiler \emph{preserves} a non-interference property if, after compilation
from a source to a target language, the property still holds.
In Patrignani and Guarnieri's framework, the source language describes
sequential execution while the target language has speculative semantics,
making their notion of compiler preservation very similar to enforcing relative NI.
}

\subsection{Execution models}
\label{subsec:predictors}


To reason about Spectre attacks, a semantics must be able to reason about the
leakage of sensitive data in a speculative \emph{execution model}.
A speculative execution model is what differentiates a speculative semantics
from standard sequential analysis,
and determines what speculation the abstract processor can perform.
For developers, choosing a proper execution model is a tradeoff:
%
On the one hand, the choice of behaviors their model allows---\ie which
microarchitectural predictors they include---determines which Spectre
variants their tools can capture.
On the other hand,
considering additional kinds of mispredictions
inevitably makes their analysis more complex.

\mypara{Spectre variants and predictors}
%
%
Most semantics and tools in \Cref{tab:mechanics} only consider the
conditional branch predictor, and thus only Spectre-PHT attacks.
%
(Mis)predictions from the conditional branch predictor are
constrained---there are only two possible choices for every decision---so
the analysis remains fairly tractable. 
Jasmin~\cite{jasmin}, Binsec/Haunted~\cite{hauntedrel},
Pitchfork~\cite{cauligi2020foundations}, and Kaibyo~\cite{cats-vs-spectre} all additionally
model \emph{store-to-load} (STL) predictions, where a processor forwards
data to a memory load from a prior store to the same address.
If there are multiple pending stores to that address, the processor may
choose the wrong store to forward the data---this is the root of a
Spectre-STL attack.
STL predictions are less constrained than predictions from the conditional
branch predictor: In the absence of additional constraints, they allow for a
load to draw data from any prior store to the same address.
%

\NEW{
%
Other control-flow mechanisms are significantly more complex:
%
Return instructions and indirect jumps can be \emph{speculatively hijacked}
to send execution to arbitrary (attacker-controlled) points in the program.%
\footnote{Including, on x86-family processors, into the \emph{middle} of an instruction~\cite{specrop}.}
An attacker can trivially hijack a victim program if they can control (mis)prediction
of the RSB (for returns)~\cite{Koruyeh2018spectre5,Maisuradze2018spectre5} or BTB (for indirect jumps)~\cite{Kocher2019spectre}.
Even without this ability, an attacker can hijack control-flow if they
speculatively overwrite the target address of a return or jump
(\eg by exploiting a prior PHT misprediction)~\cite{kiriansky2018speculative,sternberger2018spectre,spear}.
Formally, these attacks still fit within our non-interference framework---if a program
can be arbitrarily hijacked, then it will be unable to satisfy any non-interference property.
However, to formally verify that this is the case, a semantics must
model these behaviors.
}

Although capturing all speculative behaviors in a semantics is possible, the resulting
analysis is neither practical nor useful; in practice, developers need to make
tradeoffs.
For example, the semantics proposed by Cauligi \etal~\cite{cauligi2020foundations} can simulate all
of the aforementioned speculative attacks, but their analysis tool Pitchfork
only detects PHT- and STL-based vulnerabilities.
\NEW{
  On the other hand, tools like oo7 (with the ``v1.1'' pattern)~\cite{oo7} and SpecTaint~\cite{spectaint} conservatively
assume that writes to transient addresses can overwrite \emph{anything}, and thus immediately
flag this behavior as vulnerable.
}


%
The InSpectre semantics~\cite{guanciale2020inspectre} proceeds in the opposite direction---it allows
the processor to predict arbitrary values, even the values of constants.
InSpectre also allows more out-of-order behavior than most other
semantics (see \Cref{subsec:ooo})---in particular, it allows the processor to
commit writes to memory out-of-order. 
As a result, InSpectre is very expressive: It is
capable of describing a wide variety of Spectre variants both
known and unrealized.
But,
as a result, InSpectre cannot feasibly
be used to verify programs; instead, the authors pose InSpectre
as a framework for reasoning about and analyzing microarchitectural features themselves.


\mypara{Speculation windows}
Several semantics and tools in \Cref{tab:mechanics} limit
speculative execution by way of a \emph{speculation window}.
This models how hardware has finite resources for speculation,
and can only speculate through a certain number of instructions or branches
at a time.

Explicitly modeling a speculation window serves two purposes for detection
tools.
One, it reduces false positives: a mispredicted branch
will not lead to a speculative leak thousands of instructions later.
Two, it bounds the complexity of the semantics and thus the analysis.
Since the abstract processor can only speculate up to a certain depth, an
analysis tool
need only consider the latest window of instructions under speculative execution.
Some semantics refine this idea even further: Binsec/Haunted~\cite{hauntedrel}, for example,
uses different speculation windows for load-store forwarding than it uses for
branch speculation.

Speculation windows are also valuable for mitigation tools:
although tools like Blade~\cite{vassena2021} and Jasmin~\cite{jasmin} are able
to prove security without reasoning about speculation windows, modeling a
speculation window reduces the number of fences (or other mitigations)
these tools need to insert, improving the performance of the compiled code.

\mypara{Eliminating variants}
%
%
Instead of modeling all speculative behaviors, compilers and
mitigation tools can
use clever techniques to sidestep particularly problematic Spectre variants.
For example, even though Jasmin~\cite{jasmin} does not model the RSB, Jasmin
programs do not suffer from Spectre-RSB attacks: The Jasmin compiler inlines
all functions, so there are no returns to mispredict.
Mitigation tools can also disable certain classes of speculation
with hardware flags~\cite{ssbd}.
After eliminating complex or otherwise troublesome speculative behavior,
a tool need only consider those that remain.
%

\mypara{Cross-address-space attacks}
\label{marker:in-out-place} 
%
Previous systematizations of Spectre attacks~\cite{Canella2019} differentiate
between \NEW{\emph{same-address-space} and \emph{cross-address-space}} attacks.
\NEW{Same-address-space attacks} rely on
repeatedly causing the victim code to execute in order to train a
microarchitectural predictor.
\NEW{Cross-address-space attacks} are more powerful, as they allow an attacker
to perform the training step on a branch within the attacker's own code.

Most of the semantics and tools in \Cref{tab:mechanics} make no distinction
between \NEW{same-address-space and cross-address-space attacks}, as they ignore
the mechanics of training and consider all predictions to be potentially
malicious.
A notable exception is oo7~\cite{oo7}, which explicitly tracks \emph{attacker influence}.
Specifically, oo7 only considers mispredictions for
conditional branches which can be influenced by attacker input.
Thus, oo7 effectively models only \NEW{same-address-space attacks}.
Unfortunately, as a result, oo7 misses Spectre vulnerabilities in real code,
as demonstrated by Wang \etal~\cite{kleespectre}.



\subsection{Nondeterminism}
\label{subsec:proving-props}
\label{subsec:nondeterminism}

Speculative execution is inherently
\emph{nondeterministic}: Any given branch in a program may
proceed either correctly or incorrectly, regardless of the actual condition value.
More generally, speculative hijack attacks can send execution to entirely indeterminate locations.
All of the semantics in \Cref{tab:mechanics} allow these nondeterministic
choices to be actively adversarial---%
for instance, given by attacker-specified
directives~\cite{cauligi2020foundations,vassena2021}
or by consulting an abstract
oracle~\cite{cheang2019,guarnieri2020spectector,guarnieri2021contracts,mcilroy2019}.
%
These semantics
all (conservatively) assume that the attacker has full control
of microarchitectural prediction and scheduling; we explore
the different techniques they use to verify or enforce security
in the face of adversarial nondeterminism.
%

\mypara{Exploring nondeterminism}
Several Spectre analysis tools are built on some form of abstract execution:
They simulate speculative execution of the program
by tracking ranges or properties of different values.
By checking these properties throughout the program, these tools
determine if sensitive data can be leaked.
Standard tools for (non-speculative) abstract execution are
designed only to consider concrete execution paths; they must
be adapted to handle the many possible nondeterministic execution paths from speculation.
SpecuSym~\cite{specusym}, KLEESpectre~\cite{kleespectre}, and
AISE~\cite{wu-wang} handle this nondeterminism by following an
\emph{always-mispredict} strategy.
When they encounter a conditional branch, they first explore the
execution path which mispredicts this branch, up to a given speculation
depth.
Then, when they exhaust this path, they return to the correct branch.
This technique, though, only handles the conditional branch
predictor; \ie Spectre-PHT attacks.
Pitchfork~\cite{cauligi2020foundations} and Binsec/Haunted~\cite{hauntedrel}
adapt the \emph{always-mispredict} strategy to account for
out-of-order execution and Spectre-STL.
%
%
Although
it may not be immediately clear that \emph{always-mispredict}
strategies are sufficient to prove security---especially when the attacker can
make any number of antagonistic choices---%
these strategies do indeed form a sound analysis%
~\cite{guarnieri2020spectector,cauligi2020foundations,hauntedrel}.

Unfortunately,
simulating execution only works for semantics where nondeterminism is
relatively constrained:
Conditional branches are a simple boolean choice, and store-to-load predictions
are limited by the speculation window.
If we pursue other Spectre variants, we will quickly become overwhelmed---%
again, an unconstrained hijack gadget can redirect control to almost anywhere in a program.
The \emph{always-mispredict} strategy here is nonsensical at best;
abstract execution is thus necessarily limited in what it can soundly explore.

\mypara{Abstracting out nondeterminism}
Mitigation tools have more flexibility dealing with nondeterminism:
Tools like Blade~\cite{vassena2021} and oo7~\cite{oo7} apply
dataflow analysis to determine which values may be leaked along \emph{any} path,
instead of reasoning about each path individually.
Then, these tools insert speculation barriers to preemptively block potential
leaks of sensitive data.
%
%
This style of analysis comes at the cost of some precision:
Blade, for example, conservatively treats \emph{all} memory accesses as if
they may speculatively load sensitive values, as its analysis cannot reason
about the contents of memory.
\NEW{
Similarly, oo7's ``v1.1'' pattern detection conservatively flags all
(attacker-controlled) transient \emph{stores}, as they may lead to speculative hijack.
}
However, Blade and oo7---and mitigation tools in general---can afford to be less precise
than verification or detection tools; these, conversely, must
maintain higher precision to avoid floods of false positives.


\mypara{Restricting nondeterminism}
Compilers such as Swivel~\cite{swivel}, Venkman~\cite{venkman}, and
ELFbac~\cite{jenkins2020ghostbusting}
restructure programs entirely, imposing their own restricted set
of speculative behavior at the software layer.
ELFbac allocates sensitive data in separate memory regions and uses
page permission bits to disallow untrusted code from accessing
these regions---regardless of how a program may misspeculate, it
will not be able to read (and thus cannot leak) sensitive data.
Swivel 
and Venkman
compile code into carefully aligned blocks so that control flow always land at the tops
of protected code blocks, even speculatively; Swivel accomplishes
this by clearing the BTB state after untrusted execution, while Venkman
recompiles all programs on the system to mask addresses before jumping.
\NEW{
Both systems also enforce speculative control-flow integrity (CFI) checks to prevent
speculative hijacking, whether by relying on hardware features~\cite{intel-sdm}
or by implementing custom CFI checks with branchless assembly instructions.
}
Developers that use these compilers can then reason about their programs much
more simply, as the set of speculative behaviors is restricted enough to make
the analysis tractable.
Of the techniques discussed in this section,
this line of work seems the most promising:
It produces mitigation tools with strong security guarantees,
without relying on an abundance of speculation barriers (as often
results from dataflow analysis) or resorting to heavyweight
simulation (\eg symbolic execution).

\NEW{
\mypara{Open problems: Rigorous performance comparison}
To the best of our knowledge, no work has rigorously compared the performance of
all the tools in \Cref{tab:mechanics}.
Perhaps the most complete comparison is by Daniel \etal~\cite{hauntedrel},
who compare the detection tools KLEESpectre, Pitchfork, and Binsec/Haunted in terms of the
analysis time required to detect known violations in a few chosen targets.
%
%
A general and objective performance comparison is difficult, if not impossible:
The tools in \Cref{tab:mechanics} operate on different types of programs
(general-purpose, cryptographic, sandboxing) and different languages (x86, LLVM,
WebAssembly).
They also provide different security guarantees, as we discuss above.
An intermediate step towards an expanded performance comparison,
which would be a valuable contribution on its
own, would be to develop a larger corpus of known attacks on realistic
(medium-to-large-size) programs.
This corpus would help evaluate both the security and performance of existing or
newly-proposed tools.
}

\subsection{Higher-level abstractions}
\label{subsec:low-vs-high}

Spectre attacks---and speculative execution---fundamentally break our
intuitive assumptions about how programs should execute.
Higher-level guarantees about programs no longer apply:
Type systems or module systems are meaningless when even basic control flow
can go awry.
In order to rebuild higher-level security guarantees, we first need to
repair our model of how programs execute, starting from low-level semantics.
%
%
Once these foundations are firmly in place, only then can we
rebuild higher-level abstractions.

\mypara{Semantics for assembly or IRs}
The majority of formal semantics in \Cref{tab:mechanics} operate on
abstract assembly-like languages, with commands that map to simple
architectural instructions.
Semantics at this level implement control flow directly in terms
of jumps to \emph{program points}---usually indices into memory or an
array of program instructions---and treat memory as largely unstructured.
Since these low-level semantics closely correspond to the behavior of real
hardware, they capture speculative behaviors in a straightforward manner,
and provide a foundational model for higher-level reasoning.
Similarly, many concrete
analysis tools for constant-time or Spectre operate directly on
binaries or
compiler intermediate representations (IRs)%
~\cite{cauligi2020foundations,hauntedrel,binsecrel,guarnieri2020spectector,kleespectre}.
These tools operate at this lowest level
so that their analysis will be valid for the program unaltered---%
compiler optimizations for higher-level languages can end up transforming programs in
insecure ways~\cite{barthe2018,hauntedrel,binsecrel}.
As a result however, these tools necessarily
lose access to higher-level information such as control flow structure
or how variables are mapped in memory.

\mypara{Semantics for structured languages}
The semantics proposed by Jasmin~\cite{jasmin}, Patrignani and Guarnieri~\cite{patrignani2020}, and Blade~\cite{vassena2021}
build on top of these lower-level ideas to describe what we term
``medium-level'' languages---those with structured control flow and memory, \eg explicit loops
and arrays.
For these medium-level semantics, it is less straightforward to express
speculative behavior: For instance, instead of modeling speculation directly, Vassena
\etal~\cite{vassena2021} first translate programs in their source language to
lower-level commands, then apply speculative execution at that lower level.

In exchange, the structure in a medium-level semantics
lends itself well to program analysis.
For example, Vassena \etal are able to use a simple type system to prove security
properties about a program.
Barthe~\etal~\cite{jasmin} also take advantage of structured semantics:
They prove that if a sequentially
constant-time program is \emph{speculatively (memory) safe}---\ie all
memory operations are in-bounds array accesses---then the program is
also speculatively constant-time.
Since their source semantics only accesses memory through array operations,
they can statically verify whether a program is speculatively safe---and
thus speculatively secure.
%
%
An interesting question for future work is whether their concept of
speculative (memory) safety combines with other sequential security
properties to give corresponding guarantees, such as for
sandboxing, information flow, or rich type systems.

\mypara{Weak-memory-style semantics}
Weak-memory-style semantics
present a fundamentally different approach, lifting
the concept of speculative execution directly to a higher level.
As these models are abstracted away from microarchitectural details, they are
well-suited for analyzing Spectre variants in terms of data flow:
Indeed, both Colvin and Winter~\cite{colvin2019} and Disselkoen
\etal~\cite{disselkoen2019} treat Spectre-PHT as a constrained form of
instruction reordering, while Ponce de Le\'on and Kinder~\cite{cats-vs-spectre}
analyze dependency relations between instructions.

However, it remains challenging to translate a flexible semantics of this style
into a concrete analysis tool: Of the three works discussed here, only Ponce de
Le\'on and Kinder present a tool which can automatically
perform a security analysis of a target program,%
\footnote{Colvin and Winter do present a tool, but
it is only used to mechanically explore manually translated programs.}
though even they admit that it is slower than comparative tools based on operational semantics.
That said, this high-level approach to speculative semantics is certainly
underexplored compared to the larger body of work on operational semantics, and
is worthy of further investigation.

\mypara{Compiler mitigations}
With adequate foundations in place,
one avenue to regaining higher-level abstractions is to modify
compilers of higher-level languages to produce speculatively secure low-level programs.
Many compilers already include options to conservatively insert speculation barriers or
hardening into programs, which (when done properly) provides
strong security guarantees.
Although some such hardening passes have been
verified~\cite{patrignani2020}, they are overly conservative and incur a
significant performance cost.
Other compiler mitigations been shown unsound~\cite{specfuzz}---or worse, even introduce new
Spectre vulnerabilities~\cite{hauntedrel}---further reinforcing
that these techniques must be grounded in a formal semantics.
%

\mypara{Open problems: Formalization of new compilation techniques}
Swivel~\cite{swivel}, Venkman~\cite{venkman}, and
ELFbac~\cite{jenkins2020ghostbusting}
show how
the structure of code itself can provide
security guarantees at a reduced performance cost.
%
%
For instance, both Venkman and Swivel demonstrate that organizing instructions into
\emph{bundles} or \emph{linear blocks} (respectively) can
mitigate speculative hijacks, making these transient attacks
tractable to analyze and prevent.
However, none of these compiler-based approaches are yet grounded in a
formal semantics.
Formalizing these systems would increase our confidence in the strong
guarantees they claim to provide.

\mypara{Open problems: New languages}
Another promising approach is to design new languages which are inherently
safe from Spectre attacks.
Prior work has produced secure languages like FaCT~\cite{fact}, which is
(sequentially) constant-time by construction.
An extension of FaCT, or a new language built on its ideas, could prevent
Spectre attacks as well.
Vassena \etal~\cite{vassena2021} have already taken a first step in this
direction: They construct a simple \emph{while}-language which is guaranteed safe
from Spectre-PHT attacks when compiled with their fence insertion algorithm.
%
%
It would be valuable to extend this further, both to more realistic
(higher-level) languages, and to more Spectre variants.
The key question is whether dedicated language support can
provide a path to secure code that outperforms the de-facto
approach---that is, compiling standard C code and inserting Spectre mitigations.



\subsection{Expressivity and microarchitectural features}
\label{sec:expressivity}


One theme of this paper is that
a good (practical) semantics needs to have an appropriate amount of
\emph{expressivity}:
On one hand, we want a semantics which is expressive---able to
model a wide range of possible behaviors (\eg Spectre variants).
This allows us to model powerful attackers.
On the other hand, a semantics which allows too many
possible behaviors makes many analyses intractable.
%
Indeed, a fundamental purpose of semantics is to provide a reasonable abstraction
or simplification of hardware to ease analysis; a semantics which is
too expressive simply punts this problem to the analysis writer.
Thus, choosing how much expressivity to include in a semantics represents an
interesting tradeoff.

By far the most important choice for the expressivity of a semantics is which
misprediction behaviors to allow---\ie which Spectre variants to reason
about
(discussed in \Cref{subsec:predictors}).
But beyond speculative execution itself, there are many other
microarchitectural features which are relevant for a security analysis,
and which have been---or could be---modeled in a speculative semantics.
These features also affect the expressivity of the semantics, which means
that choosing whether to include them results in similar tradeoffs.

\mypara{Out-of-order execution}
\label{subsec:ooo}
Many speculative semantics simulate a processor feature called
\emph{out-of-order execution}: They allow instructions to be executed in any
order, as long as those instructions' dependencies (operands) are ready.
%
%
Out-of-order execution is mostly orthogonal to speculative execution;
in fact, out-of-order execution is not required to model
Spectre-PHT, -BTB, or -RSB---speculative execution alone is sufficient.
%
%
However, out-of-order execution is included in most modern processors, and
for that reason,\footnote{Or perhaps, because out-of-order execution
is often discussed alongside (or even confused with) speculative execution.}
many speculative semantics also model it.
Modeling out-of-order execution may provide an easier or more elegant way to
express a variety of Spectre attacks, as opposed to modeling speculative
execution alone.
%
%
Furthermore, Disselkoen \etal~\cite{disselkoen2019} and Guanciale \etal~\cite{guanciale2020inspectre}
demonstrate how to abuse out-of-order execution to conduct (at least theoretical)
novel side-channel attacks.\footnote{Disselkoen \etal~\cite{disselkoen2019}
propose to abuse compile-time instruction reordering, which is
different from microarchitectural out-of-order execution, but related.}

Although modeling out-of-order execution might make a semantics
simpler, the additional expressivity makes the resulting analysis
more complex.
Fully modeling out-of-order execution leads to an explosion in
the number of possible executions of a program; naively incorporating
out-of-order execution into a detection or mitigation tool results in an
intractable analysis.
Indeed, while Guarnieri \etal~\cite{guarnieri2021contracts} and Colvin and
Winter~\cite{colvin2019} present analysis tools based on their respective
out-of-order semantics, they only analyze very simple Spectre gadgets and not
code used in real programs.
Instead, for analysis tools based on out-of-order semantics to scale to real
programs, developers need to use lemmas to reduce the number of possibilities
the analysis needs to consider.
As one example, Pitchfork~\cite{cauligi2020foundations} operates on a set of
``worst-case schedules'' which represent a small subset of all possible
out-of-order schedules---%
the developers formally show that this reduction does not affect the
soundness of Pitchfork's analysis.


\mypara{Caches and TLBs}
Some speculative semantics and tools~\cite{specusym,mcilroy2019,kleespectre,wu-wang} include abstract models of
caches, tracking which addresses may be in the cache at
a given time.
One could imagine also including detailed models of TLBs.
As discussed in \Cref{subsec:leakage}, modeling caches or TLBs is probably
not helpful, at least for mitigation or verification tools---not only does it
make the semantics more complicated, but it potentially leads to non-portable
guarantees.
In particular, including a model of the cache usually leads to the
\contract{}{cache} leakage model, rather than the \contract{}{ct} or
\contract{}{arch} leakage models which provide stronger defensive guarantees.
Following in the tradition of constant-time programming in the
non-speculative world, it seems wiser for our analyses and mitigations to be
based on microarchitecture-agnostic principles as much as possible, and not
depend on details of the cache or TLB structure.

\mypara{Other leakage channels}
%
There are a variety of specific microarchitectural mechanisms which can
result in leakages beyond the ones we directly focus on in this paper.
For instance, in the presence of multithreading, port contention in the
processor's execution units can reveal sensitive
information~\cite{smotherspectre}; and many processor instructions, \eg
floating-point or SIMD instructions, can reveal information about their
operands through timing side channels~\cite{andrysco2015subnormal}.
Most existing semantics do not model these specific effects.
However, the commonly-used \contract{}{ct} and \contract{}{arch} leakage
models are already strong enough to capture leakages from most of these
sources:
For instance, port contention can only reveal sensitive data if the sensitive
data influenced which instructions are being executed---and the
\contract{}{ct} leakage model already considers the sensitive
data to be leaked once it influences control flow.
For variable-time instructions, most definitions of \contract{}{ct}
do not capture this leakage---but extending those definitions is
straightforward~\cite{ctverif}.
%
In both of these examples, the \contract{}{arch} leakage model captures
all leaks, as it (even more conservatively) already
considers the sensitive data as leaked once it reaches a register---long before the
data can influence control-flow or be used in an instruction.
Although modeling any of these effects more precisely can increase the
precision with which an analysis detects potential vulnerabilities, the
tradeoff in analysis complexity is probably not worth it, and for mitigation
and verification tools, the \contract{}{ct} and \contract{}{arch} leakage
models provide stronger and more generalizable guarantees.

In a similar vein, most semantics and tools do not explicitly model parallelism
or concurrency: They reason only about single-threaded programs and processors.
Instead, they abstract away these details by giving attackers broad powers in
their models---\eg complete power over all microarchitectural predictions,
and the capability to observe the full cache state after every execution
step.
The notable exceptions are the weak-memory-style semantics~\cite{colvin2019,disselkoen2019,cats-vs-spectre}---%
multiple threads are an inherent feature for
this style, making them a
promising vehicle for further exploring the
interaction between speculation and concurrency.
%

%

\mypara{Open problems: Process isolation}
In practice, a common response to Spectre attacks has been to move all secret
data into a separate process---\eg Chrome isolates different \emph{sites} in
separate processes~\cite{site-isolation}.
This shifts the burden from application and runtime system
engineers to OS engineers.
Developing Spectre foundations to model the process abstraction will
elucidate the security guarantees of such systems.
This is especially useful, as the
process boundary does not keep an attacker from performing out-of-place
training of the conditional branch predictor, nor from leaking secrets via the
cache state~\cite{Canella2019}.

\section{Related Work}

Both in industry and in academia,
there has been a lot of interest in Spectre and other transient execution
attacks.
We discuss other systematization papers that address Spectre attacks and
defenses, and we briefly survey related work which otherwise falls outside
the scope of this paper.

\subsection{Systematization of Spectre attacks and defenses}
\label{subsec:rel-soks}
Canella \etal~\cite{Canella2019} present a comprehensive systematization and
analysis of Spectre and Meltdown attacks and defenses.
They first classify transient execution attacks by whether they are a result of misprediction (Spectre) or
an execution fault (Meltdown); and further classify the attacks by
their root microarchitectural cause, yielding the nomenclature we use in this
paper (\eg Spectre-PHT is named for the \emph{Pattern History Table}).
They then categorize previously known Spectre attacks, revealing several new variants and exploitation
techniques.
%
%
Canella \etal also propose a sequence of ``phases'' for a successful Spectre or Meltdown attack, and group
published defenses by the phase they target.
A followup survey by Canella \etal~\cite{canella2020evolution} expands on the idea of attack phases,
categorizing both hardware and software Spectre defenses according to which attack phase they prevent:
Preparation, misspeculation, data access, data encoding, leakage, or decoding.
\NEW{Separately, Xiong \etal~\cite{xiongsurvey} also survey transient execution
attacks, with a specific focus on the mechanics of exploits for these attacks.}
In contrast, our systematization focuses on the formal semantics behind Spectre analysis and mitigation tools
rather than the specifics of attack variants or types of defenses.

\subsection{Hardware-based Spectre defenses}
\label{subsec:rel-hwdefs}
In this paper, we focus only on software-based techniques for existing hardware.
The research community has also proposed several hardware-based
Spectre defenses based on cache partitioning~\cite{kiriansky2018dawg}, cleaning up
the cache state after misprediction~\cite{saileshwar2019cleanupspec}, or
making the cache invisible to speculation by incorporating some separate internal
state~\cite{yan2018invisispec,ainsworth2019muontrap,khasawneh2019safespec}.
Unfortunately, attackers can still use side channels other than the cache
to exploit speculative execution~\cite{smotherspectre,schwarz2019netspectre}.
NDA~\cite{weisse2019nda}, \NEW{DOLMA~\cite{dolma}}, and Speculative Taint Tracking
(STT)~\cite{yu2019speculative} block additional speculative covert channels
by analyzing and classifying instructions that can leak information.

Fadiheh \etal~\cite{fadiheh2020} define a property for hardware execution
that they term UPEC: A hardware that satisfies UPEC will not
leak speculatively anything more than it would leak sequentially.
In other words, UPEC is equivalent to the relative non-interference property
$\nipi{\pi}{\contract{seq}{arch} \implies \contract{pht}{arch}}$.

\NEW{
The insights and recommendations from our work can guide future
hardware mitigations; properties like \contract{}{ct} or \contract{}{arch} can
serve as contracts of what software expects from hardware~\cite{guarnieri2021contracts}.
}



\subsection{Software-hardware co-design}
\label{subsec:rel-swhw}
Although hardware-only approaches are promising for future designs, they
require significant modifications and introduce non-negligible performance
overhead for all workloads.
Several works instead propose a software-hardware co-design approach.
Taram \etal~\cite{taram2019context} propose context-sensitive fencing, making
various speculative barriers available to software.
Li \etal~\cite{li2019conditional} propose memory instructions with a
conditional speculation flag.
Context~\cite{schwarz2020context} and
SpectreGuard~\cite{fustos2019spectreguard} allow software to mark secrets in
memory. This information is propagated through the microarchitecture to block
speculative access to the marked regions.
SpecCFI~\cite{koruyeh2019speccfi} suggests a hardware extension similar to
Intel CET~\cite{intel-sdm} that provides target label
instructions with speculative guarantees.
Finally, several recent proposals allow partitioning branch predictors based
on context provided by the
software~\cite{vougioukas2019brb,zhao2020lightweight}.
As these approaches require both software and hardware changes, should develop
a formal semantics to apply them correctly.

\subsection{Other transient execution attacks}
\label{sec:otherattacks}
We focus exclusively on Spectre, as other transient
execution attacks are better addressed in hardware.
For completeness, we briefly discuss these other attacks.

\mypara{Meltdown variants}
The Meltdown attack~\cite{meltdown} bypasses implicit memory permission
checks within the CPU during transient execution.
Unlike Spectre, Meltdown does not rely on executing instructions in the
victim domain, so it cannot be mitigated purely by changes to the victim's
code.
Foreshadow~\cite{Vanbulck2018} and microarchitectural data sampling
(MDS)~\cite{Canella2019Fallout,Intel2019MDS} demonstrate that transient
faults and microcode assists can still leak data from
other security domains, even on CPUs that are resistant to Meltdown.
Researchers have extensively evaluated these Meltdown-style attacks leading to new vulnerabilities~\cite{moghimi2020data,moghimi2020medusa,Schwarz2019ZL}, but most recent Intel CPUs have hardware-level mitigations for all these vulnerabilities in the form of microcode patches or proprietary hardware fixes~\cite{intelMitigationList}.

\mypara{Load value injection}
Load value injection (LVI)~\cite{vanbulck2020lvi} exploits the same root
cause as Meltdown, Foreshadow, and MDS,
but reverses these attacks: The attacker induces the transient fault into
the victim domain instead of crafting arbitrary gadgets in their own code
space.
This inverse effect is subject to an exploitation technique similar to
Spectre-BTB for transiently hijacking control flow.
Although there are software-based mitigations proposed against
LVI~\cite{vanbulck2020lvi,lvioptim}, Intel only suggests applying them to
legacy enclave software.
Like Meltdown, LVI does not need software-based mitigation on recent Intel
CPUs.

\section{Conclusion}





Spectre attacks break the abstractions afforded to us by conventional
execution models, fundamentally changing how we must reason about
security.
We systematize the community's work towards rebuilding foundations
for formal analysis atop the loose earth of speculative execution,
evaluating current efforts in a shared formal framework
and pointing out open areas for future work in this field.

We find that, as with previous work in the sequential domain, solid
foundations for speculative analyses require proper choices for
semantics and attacker models.
Most importantly, developers must consider leakage models no weaker than
\contract{}{arch} or \contract{}{ct}. Weaker models---those that only capture
leaks via memory or the data cache---lead to weaker security guarantees with no
clear benefit.
Next,
though many frameworks focus on Spectre-PHT, sound tools must consider all
Spectre variants.
Although this increases the complexity of analysis, developers can
combine analyses with structured compilation techniques to restrict or
remove entire categories of Spectre attacks by construction.
Finally, we recommend \emph{against} modeling unnecessary (micro)architectural details in
favor of the simpler \contract{}{arch} and \contract{}{ct} models; details like
cache structures or port contention introduce complexity and
reduce~portability.

When properly rooted in formal guarantees, software Spectre defenses provide a
firm foundation on which to rebuild secure systems.
We intend this systematization to serve as a reference and guide for those
seeking to build or employ formal frameworks and to develop sound Spectre defenses
with strong, precise security guarantees.

\section*{Acknowledgements}

We thank the anonymous reviewers for their insightful feedback.
We thank Matthew Kolosick for helping us understand some of the formal systems
discussed and in organizing the paper.
This work was supported in part by gifts from Intel and Google; by the NSF
under Grant Numbers CNS-2120642, CCF-1918573 and CAREER CNS-2048262; by the
CONIX Research Center, one of six centers in JUMP, a Semiconductor Research
Corporation (SRC) program sponsored by DARPA; and, by the Office of Naval
Research (ONR) under project N00014-15-1-2750.

\bibliographystyle{abbrv}
{\small
\bibliography{bib}
}

\end{document}